\documentclass[twocolumn,showpacs,showkeys,preprintnumbers,prd,superscriptaddress,nofootinbib]{revtex4-1}

\usepackage{amsmath}
\usepackage{amsfonts}
\usepackage{amssymb}
\usepackage{graphicx}
\usepackage{color}
\usepackage{hyperref}
\usepackage{booktabs}
\usepackage{hyperref}
\usepackage{cleveref}
\usepackage{color}

\Crefname{equation}{Eq.}{Eqs.}
\Crefname{figure}{Fig.}{Figs.}
\Crefname{section}{Sec.}{Secs.}

\usepackage{etoolbox}
\makeatletter
\appto{\appendix}{%
  \@ifstar{\def\theequation@prefix{A.}}%
          {}%
}
\makeatother

\begin{document}

\title{Effective field description of the Anton-Schmidt cosmic fluid}

\author{Salvatore Capozziello}
\email{capozzie@na.infn.it}
\affiliation{Dipartimento di Fisica, Universit\`a di Napoli  ``Federico II'', Via Cinthia, I-80126, Napoli, Italy.}
\affiliation{Istituto Nazionale di Fisica Nucleare (INFN), Sez. di Napoli, Via Cinthia 9, I-80126 Napoli, Italy.}
\affiliation{Gran Sasso Science Institute, Via F. Crispi 7, I-67100, L' Aquila, Italy.}

\author{Rocco D'Agostino}
\email{rocco.dagostino@roma2.infn.it}
\affiliation{Dipartimento di Fisica, Universit\`a degli Studi di Roma ``Tor Vergata'', Via della Ricerca Scientifica 1, I-00133, Roma, Italy.}
\affiliation{Istituto Nazionale di Fisica Nucleare (INFN), Sez. di Roma ``Tor Vergata'', Via della Ricerca Scientifica 1, I-00133, Roma, Italy.}

\author{Roberto Giamb\`o}
\email{roberto.giambo@unicam.it}
\affiliation{Scuola di Scienze e Tecnologie, Universit\`a di Camerino, I-62032, Camerino, Italy.}

\author{Orlando Luongo}	
\email{orlando.luongo@lnf.infn.it}
\affiliation{Scuola di Scienze e Tecnologie, Universit\`a di Camerino, I-62032, Camerino, Italy.}
\affiliation{Istituto Nazionale di Fisica Nucleare, Laboratori Nazionali di Frascati, 00044 Frascati, Italy.}
\affiliation{Instituto de Ciencias Nucleares, Universidad Nacional Aut́onoma de Ḿexico, AP 70543, Mexico, DF 04510, Mexico.}


\begin{abstract}
The effective theory of the Anton-Schmidt cosmic fluid within the Debye approximation is investigated. In this picture,  the universe is modeled out by means of a medium without cosmological constant. In particular, the Anton-Schmidt  representation of matter describes the pressure of crystalline solids under deformations imposed by isotropic stresses. The approach  scheme is related to the fact that  the universe deforms under the action of the cosmic expansion itself. Thus, we frame the dark energy term as a function of scalar fields and  obtain the corresponding dark energy potential $V(\varphi)$. Different epochs of the universe evolution are investigated in terms of the evolution of $\varphi$. We show how the Anton-Schmidt equation of state is capable of describing both late and early epochs of  cosmic evolution. Finally, numerical bounds on the Anton-Schmidt model with $n=-1$ are derived through a Markov Chain Monte Carlo analysis on the combination of data coming from type Ia Supernovae, observations of Hubble parameter and baryon acoustic oscillations. Statistical comparison with the $\Lambda$CDM model is performed by the AIC and BIC selection criteria. Results are in excellent agreement with the low-redshift data. A further generalization of the model is presented to satisfy the theoretical predictions at early-stage cosmology.
\end{abstract}



\maketitle

\section{Introduction}

Dark energy is characterized by a negative equation of state which violates the Zeldovich limit and turns out to be highly different from standard matter \cite{Supernovae,Peebles03,Weinberg13}.  Several approaches have been used to model dark energy in terms of first principles \cite{Bamba12,Padmanabhan03,Sahni06,Copeland06,review} or through modification of gravity \cite{Capozziello11,NojiriOdintsov, Schmidt07}. An interesting case is to consider standard matter with a non-vanishing pressure which provides a different equation of state depending on the stages of universe's evolution \cite{Bharadwaj03,Saxton10}. In other words, is it possible that matter passes from $\omega_m=0$ to $\omega_m<0$. To enable the process that permits matter to pass from a pressureless equation of state to a negative pressure, we consider that matter bids to Anton-Schmidt's equation of state \cite{AntonSchmidt} and satisfies the Debye approximation \cite{Debye}. This is allowed since the universe is expanding, so that the thermodynamics associated to the matter fluid changes with time and is not perfectly a thermodynamics of equilibrium. In fact, the Anton-Schmidt representation of matter describes the pressure of crystalline solids under deformations imposed by isotropic stresses.  So that, if one considers the universe to deform under the action of cosmic expansion, the equation of state becomes negative as a natural consequence of its functional form \cite{solid}.

The advantage of the Anton-Schmidt description is that one has a non-vanishing and physically-supported pressure defined after a precise redshift domain. The cosmic acceleration is recovered as consequence of approximating matter with the Anton-Schmidt's approach. In such a picture, dark energy is featured as an isotropic medium and can be described by means of an effective scalar field description \cite{Starobinsky98,Guo07,Kamenshchik12}. This scenario has been first introduced in the field of condensed matter \cite{Li14,Beck04}.

In this paper, we show that if matter obeys Anton-Schmidt's equation of state the universe speeds up without the need of the cosmological constant. Hence, we perform an effective representation of the dark energy potential associated to Anton-Schmidt's equation of state. We describe its evolution in terms of the scale factor and we portray the limits at which Anton-Schmidt's potential stops being valid. In our analysis, we portray the effective field description in terms of unconstrained fields, employing a similar formalism of inflation \cite{Guth81,Linde82,Sebastiani15}. We write the action of the self-gravitating medium in the presence of gravity and then we give a thermodynamic interpretation of the scalar field and potential in terms of thermodynamic variables.
Relevant consequences are based on the initial settings of the scalar field and occur only as the corresponding volume takes a given value. This scenario admits, as limiting case, the approach of logotropic dark energy \cite{Chavanis15,Ferreira17}.

We demonstrate how to extend logotropic models invoking first principles based on thermodynamics of expanding media in analogy to solid state physics.
To this end, we consider the  Gr\"uneisen parameter $\gamma_G$ \cite{Gruneisen12} and we highlight its thermal and microscopic interpretations. Since its definition comes from the thermodynamics of expanding media, we show that under the quasi-harmonic approximation, it is possible to relate the macroscopic definition of $\gamma_G$ to its microscopic definition. Motivated by such a definition, we show that the Gr\"uneisen parameter can take values compatible with current universe dynamics. Hence, we investigate Anton-Schmidt's equation of state in which the pressure is exactly integrable as $\gamma_G=5/6$. Afterwards, we underline the main physical properties of this solution and finally we propose a common origin between Anton-Schmidt and logotropic dark energy.

We check the validity of our approach at late and early times. The theoretical consequences of our scenario are analyzed at the level of background cosmology. 
We underline the differences with respect to the pure logotropic models, as  testified by the adiabatic sound speed expressed in terms of the logotropic one.
Further, we constrain the free parameters of our model by means of a Monte Carlo analysis at small redshift domains, with cosmic surveys provided by Supernovae Ia, Hubble rate data and baryon acoustic oscillations measurements. In addition, we compare our numerics with the predictions of the $\Lambda$CDM model and discuss the compatibility of the results from observations at late times
with high redshift domains.

The work is organized as follows.
In \Cref{sec:FT}, we give the field theory description of the Anton-Schmidt cosmic fluid describing an accelerating universe. Specifically, we discuss the conditions under which Anton-Schmidt's equation of state can explain the cosmological dynamics at both early and late phases, and we derive the analytical form of the potential in terms of the matter fields.
In \Cref{sec:background}, we describe the features of the Anton-Schmidt model at the level of background cosmology. \Cref{sec:constraints} is dedicated to the observational constraints from low-redshift data and to the statistical comparison with the $\Lambda$CDM model.
In \Cref{sec:early}, we discuss the consequences of the obtained results at early times cosmology.
Finally, in \Cref{sec:conclusion} we summarize our findings and present the future perspectives of our work.

Throughout the paper, we use units such that $8\pi G=c=1$.


\section{The Anton-Schmidt fluid}
\label{sec:FT}

In standard general relativity the interaction between gravity and matter is described by the Einstein-Hilbert action
\begin{equation}
S=\int d^4x \sqrt{-g}\left[\dfrac{R}{2}+\mathcal{L}_m(\varphi,\dot\varphi)\right],
\end{equation}
where $g$ is the determinant of the metric $g_{\mu\nu}$ and $R$ the Ricci curvature. The matter Lagrangian $\mathcal{L}_m$ may be written in terms of scalar fields, $\varphi$. The universe is assumed to be described by a perfect fluid whose energy-momentum tensor is obtained as
\begin{equation}
T_{\mu\nu}=-\dfrac{2}{\sqrt{-g}}\dfrac{\delta \mathcal{L}_m}{\delta g^{\mu\nu}}\ .
\end{equation}

It is possible to rewrite the components of the energy-momentum in terms of scalar fields following the scheme \cite{Barrow}:
\begin{align}
\epsilon_\varphi=\dfrac{1}{2}\dot{\varphi}^2+V(\varphi)	\label{eq:rho_fi}\ ,\\
P_\varphi=\dfrac{1}{2}\dot{\varphi}^2-V(\varphi)\label{eq:p_fi}\ ,
\end{align}
where we identify $\epsilon_\varphi$ and $P_\varphi$ as the energy density and pressure of the Anton-Schmidt fluid according to the following derivation.

\subsection{A macroscopic formulation of the Anton-Schmidt fluid}

We consider the universe filled with a single fluid described by an equation of state with logarithmic-power law form. Such a prescription is analogous of crystalline solids under isotropic deformations, namely the Anton-Schmidt equation of state, which has been recently considered to frame the cosmic acceleration \cite{solid}:
\begin{equation}
P={A\left(\dfrac{\epsilon}{\epsilon_\ast}\right)}^{-n}\ln \left(\dfrac{\epsilon}{\epsilon_\ast}\right),
\label{eq:Anton-Schmidt}
\end{equation}
where $\epsilon_\ast$ is a reference energy density\footnote{The physical meaning of $\epsilon_\ast$ will be discussed in the next sections.}, and $n$ is related to the Gr\"uneisen parameter $\gamma_G$ \cite{Gruneisen12} through $n=-\frac{1}{6}-\gamma_G$. In the original formulations \cite{AntonSchmidt,solid} $\epsilon$ is replaced by the rest-mass density, $\rho$. However, we here assume $\epsilon$ to account for the total energy density of the cosmic fluid, as also examined in \cite{Odintsov18}. We are, in fact, interested in studying the late-time universe dynamics by allowing a modified equation of state for the dark energy fluid inspired by the Anton-Schmidt case.
From a theoretical point of view, the Anton-Schmidt cosmological model \cite{solid} can be described by means of a scalar field $\varphi$ and a self-interacting potential $V(\varphi)$ defining the effective Lagrangian
\begin{equation}
\mathcal{L}_m=\mathcal K(\dot \varphi) -V(\varphi)\ ,
\end{equation}
where $\mathcal K(\dot \varphi)$ is a generic kinetic term which can be recast as in the standard case, i.e. $\mathcal{L}_m=\frac{1}{2}\dot{\varphi}^2 -V(\varphi)$. In such a case, one employs the simplest assumption on $\mathcal K(\dot\varphi)$.

Following these prescriptions and the cosmological principle \cite{Peebles93}, we consider a homogeneous and isotropic flat universe described by the Friedmann-Lema{\^i}tre-Robertson-Walker (FLRW) metric \cite{Planck15}:
\begin{equation}
ds^2= dt^2-a(t)^2\left[dr^2+ r^2( d\theta^2+ \sin^2 \theta\ d\phi^2)\right],
\end{equation}
where $a(t)$ is the scale factor\footnote{Normalized by $a(t_0)=1$.}. Hence, the Friedmann equation and the continuity equation are respectively given by
\begin{align}
&H^2\equiv{\left(\dfrac{\dot{a}}{a}\right)}^2=\dfrac{\epsilon}{3} \label{eq:Friedmann}, \\
&\dot{\epsilon}+3\left(\dfrac{\dot{a}}{a}\right)(\epsilon+P)=0 \label{eq:continuity}\ ,
\end{align}
where $\epsilon$ and $P$ are, respectively, the density and the pressure for the fluid in Eq. \eqref{eq:Anton-Schmidt}.

\subsection{The Anton-Schmidt  cosmological dynamics}

The energy density of the fluid is found by plugging \Cref{eq:Anton-Schmidt} into \Cref{eq:continuity}. An immediate solution occurs for $n=-1$, which corresponds to a precise value of the so-called dimensionless Gr\"uneisen parameter $\gamma_G$ \cite{Gruneisen12}. Its meaning comes from thermodynamic properties of the material via $\gamma_G=\frac{\alpha V K_T}{C_V}$, where $\alpha$, $K_T$ are respectively the thermal coefficient and the isothermal bulk modulus and also $C_V$ the heat capacity at constant volume.

Under the quasi-harmonic approximation, the macroscopic definition becomes indistinguishable from the corresponding microscopic picture and can be easily related to the Debye temperature \cite{Debye} defined as $\theta_D=	\hbar \omega_D/k_B$, where $\hbar$ and $k_B$ are the Planck's and Boltzmann's constants, respectively, while $\omega_D$ the maximum vibrational frequency of the medium under exam. The $\gamma_G$ range typically spans into the interval $\sim 1\div2$. The limiting case $n=-1$ corresponds to isotropic and homogeneous expansion and turns out to be relevant in the case of cosmology.
In that case, integrating \Cref{eq:continuity} leads to
\begin{equation}
\epsilon=\epsilon_\ast \exp\left\{-\dfrac{\epsilon_\ast}{A}+\frac{1}{A}\left(\frac{C}{a^3}\right)^{\frac{A}{\epsilon_\ast}}\right\} ,
\label{eq:rho}
\end{equation}
where $C$ is an integration constant. Combining Eqs. \Cref{eq:rho} and \Cref{eq:Anton-Schmidt} enables to get $P$ as function of $a(t)$:
\begin{equation}
P=\left[-\epsilon_\ast+\left(\frac{C}{a^3}\right)^{\frac{A}{\epsilon_\ast}}\right]\exp\left\{-\dfrac{\epsilon_\ast}{A}+\frac{1}{A}\left(\frac{C}{a^3}\right)^{\frac{A}{\epsilon_\ast}}\right\}.
\label{eq:p}
\end{equation}
We thus distinguish two epochs:

\subsubsection{Late times $(a\gg1)$}

\noindent This scenario corresponds to a phase in which dark energy dominates over the other species. Combining \Cref{eq:rho,eq:p}, we get
\begin{equation}
P=\epsilon\left[-1+\dfrac{1}{\epsilon_\ast}{\left(\frac{C}{a^3}\right)}^\xi \right],
\label{eq:p-rho}
\end{equation}
where we have defined $\xi\equiv A/\epsilon_\ast$. In the limit of large scale factor, one has
\begin{equation}
\dfrac{1}{\epsilon_\ast}{\left(\frac{C}{a^3}\right)}^\xi \ll 1\,,
\label{big a}
\end{equation}
under the condition $\xi>0$, and therefore
\begin{equation}
P\approx -\epsilon\ .
\end{equation}
This corresponds to a de-Sitter phase capable of accelerating the universe today. Clearly, this case is not \emph{perfectly} matchable with the $\Lambda$CDM model but approximates it fairly well in the small redshift domain. Hence, the Anton-Schmidt approach provides small departures at the level of background cosmology.

\subsubsection{Early times $(a\ll1)$}
\label{subsec:early}

\noindent For small values of the scale factor, we have
\begin{equation}
{\left(\frac{C}{a^3}\right)}^\xi \gg 1\ ,
\label{small_a_1}
\end{equation}
if $\xi>0$. To recover approximatively a matter-dominated universe $(P\approx 0)$, the quantity in the square brackets of \Cref{eq:p-rho} has to be vanishingly small, \emph{i.e.}
\begin{equation}
\dfrac{1}{\epsilon_\ast} {\left(\frac{C}{a^3}\right)}^\xi  \approx 1\ .
\label{small_a_2}
\end{equation}
Combining Eqs. \eqref{small_a_1} and \eqref{small_a_2}, we obtain the condition $\epsilon_\ast \gg 1$.
In the matter-dominated era $\epsilon\sim \rho$, where $\rho$ is the mass density. Therefore, the aforementioned constraint becomes $\rho_\ast \gg 1$. In \cite{Chavanis15} $\rho_\ast$ has been identified with the Planck density. However, considerations on the linear growth rate of density perturbations show that $\rho_\ast$ is actually much larger than the Planck density \cite{Ferreira17}.

The equation of state parameter $\omega=P/\epsilon$ can be calculated by using \Cref{eq:rho,eq:p}:
\begin{equation}
\omega=-1+\dfrac{1}{\epsilon_\ast}{\left(\dfrac{C}{a^3}\right)}^\xi\ ,
\label{eq:w}
\end{equation}
which takes values in the interval $-1<\omega<0$. We note that, for very large values of the scale factor $(a\rightarrow\infty)$, $\omega\rightarrow -1$, which describes an accelerated expansion driven by the cosmological constant.

In the next paragraph, we shall derive the expressions of the physical quantities characterizing the Anton-Schmidt cosmic fluid in terms of the scalar field $\varphi$.

\section{The Anton-Schmidt effective scalar field}
\label{subsec:scalar field}

We here derive the form of $\varphi$ in terms of $a(t)$ and then we trace its evolution for Anton-Schmidt's dark energy term. To do that, we start from the kinetic energy of $\varphi$:
\begin{equation}
\dot{\varphi}^2=\left(\frac{C}{a^3}\right)^\xi\exp\left\{-\dfrac{1}{\xi}+\frac{1}{A}\left(\frac{C}{a^3}\right)^\xi\right\} .\label{eq:kinetic energy}
\end{equation}
Using \Cref{eq:Friedmann}, we obtain that \Cref{eq:kinetic energy} becomes
\begin{equation}
\dot{\varphi}^2=\dfrac{3}{\epsilon_\ast}{\left(\dfrac{\dot{a}}{a}\right)}^2 \left(\frac{C}{a^3}\right)^\xi .
\end{equation}
The scalar field as a function of the scale factor is obtained by integrating the above relation:
\begin{equation}
\varphi=\varphi_0+ \dfrac{2}{\xi\sqrt{3\epsilon_\ast}}\left(\frac{C}{a^3}\right)^{{\xi\over2}}\ .
\label{eq:fi}
\end{equation}

\begin{figure}[h!]
\begin{center}
\includegraphics[width=3.3in]{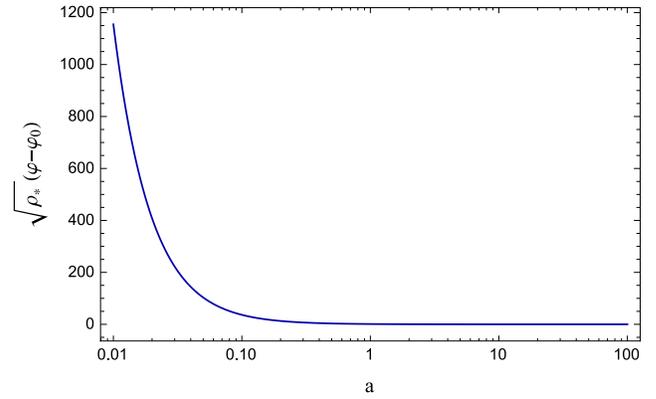}
\caption{Dynamical evolution of the scalar field for $\xi=C=1$.}
\label{fig:phi}
\end{center}
\end{figure}

\begin{figure}[h!]
\begin{center}
\includegraphics[width=3.3in]{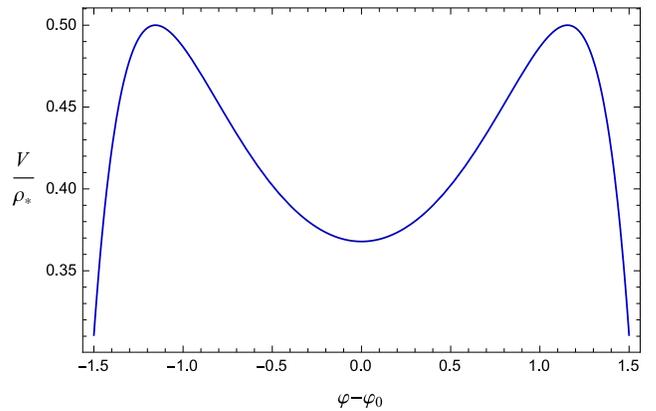}
\caption{Functional behaviour of the scalar field potential for $\xi=1$.}
\label{fig:V}
\end{center}
\end{figure}

We can calculate the potential of the scalar field by combining \Cref{eq:rho_fi,eq:p_fi} and using \Cref{eq:rho,eq:p}:
\begin{equation}
V=\left[\epsilon_\ast -\dfrac{1}{2}\left(\frac{C}{a^3}\right)^\xi\right] \exp\left\{-\dfrac{1}{\xi}+\frac{1}{A}\left(\frac{C}{a^3}\right)^\xi\right\} .
\label{eq:V}
\end{equation}
To obtain the potential in terms of the field, we invert \Cref{eq:fi} and plug it into \Cref{eq:V}. So, we finally get:
\begin{equation}
V(\varphi)=\epsilon_\ast\left[1-\dfrac{3}{8}\xi^2{(\varphi-\varphi_0)}^2\right] \exp\left\{-\dfrac{1}{\xi}+\dfrac{3}{4}\xi{(\varphi-\varphi_0)}^2\right\} .
\label{eq:V(fi)}
\end{equation}
The potential is symmetric around $\varphi_0$, where it takes the minimum value:
\begin{equation}
V_{min}=\epsilon_\ast e^{-1/\xi}\ .
\label{eq:V_min}
\end{equation}
Further, $V(\varphi)$ presents two maxima at
\begin{equation}
\varphi_{max}=\varphi_0\pm \dfrac{2}{\xi}\sqrt{\dfrac{2-\xi}{3}}\ ,
\label{eq:fi_max}
\end{equation}
where it takes the value
\begin{equation}
V_{max}=\dfrac{1}{2}\epsilon_\ast \xi\ e^{-1+1/\xi}\ .
\end{equation}
Requiring that \Cref{eq:fi_max} must be a real quantity, we obtain the following constraint:
\begin{equation}
0<\xi<2\ .
\end{equation}
In \Cref{fig:phi} and \Cref{fig:V}, we show the functional behaviours of the scalar field and the potential, respectively. As far as the constants are concerned, we adopt the indicative values of $\xi=1$ and $C=1$.

The density and pressure in terms of the field read, respectively
\begin{align}
&\epsilon_\varphi= \epsilon_\ast \exp\left\{-\dfrac{1}{\xi}+\dfrac{3}{4}\xi(\varphi-\varphi_0)^2\right\}	 , \\
&P_\varphi= \epsilon_\ast \left[-1+\dfrac{3}{4}\xi^2(\varphi-\varphi_0)^2\right] \exp\left\{-\dfrac{1}{\xi}+\dfrac{3}{4}\xi(\varphi-\varphi_0)^2\right\} ,
\end{align}
which yield to the following expression for the equation of state:
\begin{equation}
\omega_\varphi=-1+\dfrac{3}{4}\xi^2(\varphi-\varphi_0)^2.
\end{equation}
Consistently with what found before, under the condition $\xi \ll 1$, we have $\omega_\varphi\rightarrow -1$ mimicking the effect of the cosmological constant.

\section{The Anton-Schmidt background cosmology}
\label{sec:background}

The single Anton-Schmidt fluid in the Debye approximation explains the cosmic speed up without resorting to the cosmological constant \cite{solid}.
As the universe expands, Anton-Schmidt's equation of state naturally provides epochs of deceleration and other phases characterized by acceleration. This picture is true only when the universe temperature reaches a precise range of values. If the temperature increases the Anton-Schmidt approximation fails to be predictive. This landscape will be faced in the next sections when we reconsider the Anton-Schmidt behaviour at early epochs. At small redshifts, $n$ can take in principle any values. However,  the cosmological consequences are not easily matchable to early times without adding a pressureless fluid into the analysis. We here show that the case $n=-1$ is the best choice at small redshifts and deserves explanations versus the case of free $n$. A free $n$ is mostly viable as the Anton-Schmidt does not hold, leading to the initial phases of the universe evolution. We discuss later the properties of $n$ in function of the temperature $T$.

To study the cosmological features of the choice $n=-1$ at the background level, and to compare the results with the ones relative to the case of free $n$ obtained in \cite{solid}, we consider the original formulation of the Anton-Schmidt pressure
\begin{equation}
P=A\left(\dfrac{\rho}{\rho_\ast}\right)\ln \left(\dfrac{\rho}{\rho_\ast}\right) ,
\label{eq:n=-1}
\end{equation}
where $\rho$ stands for the rest-mass energy.
We assume that the FLRW universe is filled with a perfect fluid of total energy density $\epsilon$.
Under the hypothesis of adiabatic heat exchanges, the first law of thermodynamics reads
\begin{equation}
d\epsilon=\left(\dfrac{\epsilon+P}{\rho}\right)d\rho\ ,
\end{equation}
which can be integrated into
\begin{equation}
\epsilon= \rho+\rho \int^{\rho}d\rho'\dfrac{P(\rho')}{\rho'^2}\ .
\end{equation}
For $P(\rho)$ as given in \Cref{eq:n=-1}, one soon obtains
\begin{equation}
\epsilon=\rho +\dfrac{A}{2}\left(\dfrac{\rho}{\rho_*}\right)\ln^2\left(\dfrac{\rho}{\rho_*}\right) .
\label{eq:total rho}
\end{equation}
It is important to stress that the above solution cannot be recovered from the one obtained in \cite{solid} as $n\rightarrow -1$, since the case $n=-1$ is a different analytical case with respect to the free case $n\neq-1$. It follows that the Anton-Schmidt fluid with $n=-1$ represents a \emph{stand-alone scenario} worth to be studied, whereas the corresponding implications in cosmology are expect to differ from pure logotropic models examined in \cite{Chavanis17}.
As $(a\ll 1)$, the first term of \Cref{eq:total rho} dominates over the other species as in a matter-dominated universe.
Instead, as $a\gg 1$ the pressure becomes negative as in a dark energy dominated-universe. This allows one to split the total density into two contributions $(\epsilon=\epsilon_m + \epsilon_{de})$, in analogy to the standard cosmological scenario:
\begin{align}
&\epsilon_m=\dfrac{\rho_{m,0}}{a^3}\ , \label{eq:rho_m}\\
&\epsilon_{de}=\dfrac{\epsilon_{de,0} }{a^3}-\dfrac{3A}{a^3}\left(\dfrac{\rho_{m,0}}{\rho_*}\right)\ln a \ln \left(\dfrac{\rho_{m,0}}{\rho_*}a^{-3/2}\right),
\label{eq:rho_de}
\end{align}
where
\begin{equation}
\epsilon_{de,0}=\dfrac{A}{2}\left(\dfrac{\rho_{m,0}}{\rho_\ast}\right)\ln^2\left(\dfrac{\rho_{m,0}}{\rho_\ast}\right).
\label{eq:rho_de0}
\end{equation}
\Cref{eq:rho_m,eq:rho_de} correspond to matter and dark energy term, respectively. Defining the normalized density parameters as $\Omega_{m,0}\equiv \epsilon_{m,0}/\epsilon_{c,0}$ and $\Omega_{de,0}\equiv\epsilon_{de,0}/\epsilon_{c,0}=1-\Omega_{m,0}$, where $\epsilon_{c,0}\equiv3H_0^2$ is the present critical density, we can write \Cref{eq:Friedmann} as
\begin{align}
H^2&=H_0^2\left(\dfrac{\epsilon_m}{\epsilon_{c,0}}+\dfrac{\epsilon_{de}}{\epsilon_{c,0}}\right) \\
&=H_0^2\left[\dfrac{\Omega_{m,0}}{a^3}+\dfrac{\Omega_{de,0}}{a^3}\left(1-6B\ln a+9B^2 \ln^2 a\right)\right],  \nonumber
\label{eq:Hubble}
\end{align}
where we defined a new parameter:
\begin{equation}
B\equiv\ln^{-1}\left(\dfrac{\rho_{m,0}}{\rho_\ast}\right) .
\label{eq:B}
\end{equation}
It is worth noting that the above defined $B$ turns out to be quite different from the one given by logotropic models. In fact, in the case of logotropic models, it represents the dimensionless logotropic temperature defined as \cite{Chavanis17}
\begin{equation}
B_{log}\equiv \left[\ln\left(\dfrac{\rho_{\ast}}{\rho_{m,0}}\right)-1\right]^{-1}\ .
\end{equation}
If $\rho_\ast$ is identified with the Planck density as assumed in \cite{Chavanis15,Chavanis17}, one finds $0<B_{log}\ll 1$. Instead, in the present case, we expect to have $B<0$ due to the condition $\rho_\ast\gg 1$. In the following Section, we will provide observational constraints on this parameter and we will see that they differ from the bounds obtained in \cite{Chavanis17}.

Therefore, one can write the total density and pressure in terms of $B$ by:
\begin{equation}
\dfrac{\epsilon}{\epsilon_{c,0}}=\dfrac{\Omega_{m,0}}{a^3}+\dfrac{1-\Omega_{m,0}}{a^3}\left(1-6B\ln a+9B^2 \ln^2 a\right)
\end{equation}
and
\begin{align}
P&=\dfrac{A}{a^3}\left(\dfrac{\rho_{m,0}}{\rho_*}\right)\left[\ln\left(\dfrac{\rho_{m,0}}{\rho_*}\right)-3\ln a\right] 	\nonumber	\\
&=\dfrac{A}{a^3}e^{1/B}\left[\dfrac{1}{B}-3\ln a \right]\,.
\end{align}
Moreover, \Cref{eq:rho_de0} reads
\begin{equation}
\epsilon_{de,0}=\dfrac{A}{2}\dfrac{e^{1/B}}{B^2}\ ,
\end{equation}
so that
\begin{equation}
P=2\epsilon_{c,0}\dfrac{1-\Omega_{m,0}}{a^3}(B-3B^2\ln a)\ .
\label{eq:P}
\end{equation}
Thus, the total equation of state takes the form
\begin{equation}
\omega=\dfrac{2B-6B^2 \ln a}{(1-\Omega_{m,0})^{-1}-6B\ln a+9B^2 \ln^2 a}\ .
\end{equation}
Since $B<0$ and $\Omega_{m,0}<1$, $\omega$ takes negative values and increases until it reaches 0 at infinity scale factor. In particular,  at present $(a=1)$ we have
\begin{equation}
\omega_0=2B(1-\Omega_{m,0})\ .
\end{equation}
Then, identifying the total pressure of the fluid with the pressure of the dark energy term \cite{solid}, one can derive the expression for the dark energy equation of state parameter $\omega_{de}=P/\rho_{de}$:
\begin{equation}
\omega_{de}=\dfrac{2B}{1-3B\ln a}\ ,
\end{equation}
which is a negative quantity and increases until it vanishes at infinity scale factor. Today, it takes the following value:
\begin{equation}
\omega_{de,0}=2B\ .
\end{equation}
The $\Lambda$CDM paradigm is recovered as $B\rightarrow-\frac{1}{2}$ for $a=1$ only. When $a\neq1$ at small redshift one gets $\omega_{de}\approx 2B+6B^2(a-1)$. This outcome is comparable with the Chevallier-Polarski-Linder (CPL) parametrization, when the two constants, $\omega_0$ and $\omega_1$, are intertwined among them. A direct comparison with the CPL parametrization and additional models will be provided in the rest of the work.

An interesting quantity to analyze is the sound speed, which plays a crucial role for cosmological perturbations \cite{Mukhanov93}. 
In the theory of structure formation, the length above which the perturbations grow is in fact determined by the sound speed which, for an adiabatic fluid, reads
\begin{equation}
c_s^2\equiv \dfrac{\partial P}{\partial\epsilon}\ .
\label{eq:cs}
\end{equation}
For the Anton-Schmidt pressure given in \Cref{eq:n=-1}, we can use \Cref{eq:total rho} to obtain
\begin{equation}
c_s^2=\left(\dfrac{\partial P}{\partial\rho}\right)\left(\dfrac{\partial\epsilon}{\partial\rho}\right)^{-1}=\dfrac{A\left[1+\ln\left(\dfrac{\rho}{\rho_\ast}\right)\right]}{\rho_\ast+\dfrac{A}{2}\left[2+\ln\left(\dfrac{\rho}{\rho_\ast}\right)\right]\ln\left(\dfrac{\rho}{\rho_\ast}\right)}\ .
\end{equation}

In the matter epoch $\epsilon\sim \rho$ and, consequently, $c_s^2\equiv\partial P/\partial\rho$. Then, we can calculate the adiabatic sound speed for the Anton-Schmidt fluid in terms of the parameter $B$ by using
\begin{align}
&\dfrac{\partial{\rho}}{\partial{a}}=-\dfrac{3\epsilon_{c,0}\Omega_{m,0}}{a^4}\ , \\
&\dfrac{\partial P}{\partial a}=-\dfrac{6\epsilon_{c,0}(1-\Omega_{m,0})}{a^4}B(1+B-3B\ln a)\ .
\end{align}
One thus finds
\begin{equation}
c_s^2={\left(\dfrac{\partial{\rho}}{\partial{a}}\right)}^{-1}\dfrac{\partial P}{\partial a}=\dfrac{2B(1-\Omega_{m,0})(1+B-3B\ln a)}{\Omega_{m,0}}\ .
\label{eq:cs nuovo}
\end{equation}
By virtue of the aforementioned comments, due to the difference in the definition of the parameter $B$ between pure logotropic models and our paradigm, we find that
\begin{equation}\label{comparison}
c_s=\sqrt{\frac{2(1+B)(3B\ln a-B-1)}{a^3}}\ c_{s,log}\,,
\end{equation}
with $B$ given as in \Cref{eq:B} and $c_{s,log}$ being the logotropic sound speed. A dutiful caveat is that, at the level of small perturbations, $n$ becomes a function of the temperature. In such a way, we cannot conclude that $n=-1$ holds true throughout the whole universe's expansion history and the corresponding expression for the sound speed \Cref{eq:cs nuovo} may be not valid at all redshift regimes. This enables the process of structure formation and does not influence its dynamics, indicating that the Anton-Schmidt model works well even at the level of early-time cosmology \cite{solid}. We will better face this problem later in the text, as we discuss the dependence on the temperature of $n$.
We also note that our analysis does not take into account nonlinear effects, which may lead to significant modifications also at the level of background cosmology (see \cite{Avelino14} and references therein).


\section{Observational constraints}
\label{sec:constraints}

In this section, we employ cosmological data to place observational constraints on the Anton-Schmidt model in the case of $n=-1$. To do so, we combine the Supernovae Ia data of the Joint Light-curve Analysis (JLA) catalogue \cite{Betoule14}, the Observational Hubble data (OHD) acquired through the differential age method \cite{Jimenez02} and a collection of Baryon Acoustic Oscillations measurements (BAO).

\subsection{JLA Supernovae Ia}

The JLA sample \cite{Betoule14} consists of 740 type Ia Supernovae (SNe) up to redshift $z\simeq 1.3$. The catalogue provides the redshift of each SN together with its $B$-band apparent magnitude $(m_B)$, the stretch $(X_1)$ and the colour factor at maximum brightness $(C)$.  The theoretical distance modulus of a SN is defined as
\begin{equation}
\mu_{th}(z)=25+5\log_{10} d_L(z)\ ,
\end{equation}
where $d_L(z)$ is the luminosity distance given by
\begin{equation}
d_L(z)=(1+z)\int_0^z\dfrac{dz'}{H(z')}\ .
\end{equation}
Statistical analyses are done by comparing the theoretical distance modulus  with its observational form:
\begin{equation}
\mu_{obs}=m_B-(M_B-\alpha X_1+\beta C)\ .
\end{equation}
Here, $\alpha$ and $\beta$ are constant nuisance parameters, while $M_B$ is the SN absolute magnitude defined as
\begin{equation}
M_B=
\begin{cases}
 M, & \text{if}\ M_{host}<10^{10}M_{Sun}\\
  M+\Delta_M, & \text{otherwise}
 \end{cases}
\end{equation}
where $M_{host}$ is the stellar mass of the host galaxy and $\Delta_M$ is an additional nuisance parameter. Encoding the statistical and systematic uncertainties on the light-curve parameters in the covariance matrix $\textbf{C}$, one can write the normalized Likelihood as follows:
\begin{equation}
\mathcal{L}_{SN}=\dfrac{1}{{|2\pi \textbf{C}|}^{1/2}}\exp\left[-\dfrac{1}{2}\left(\mu_{th}-\mu_{obs}\right)^\dagger \textbf{C}^{-1}\left(\mu_{th}-\mu_{obs}\right)\right].
\end{equation}

\subsection{Observational Hubble data}

The differential age method \cite{Jimenez02} allows to obtain model-independent estimations of the Hubble rate. This technique is based on measuring the age difference of close passively evolving red galaxies, which are considered as cosmic chronometers. $H(z)$ measurements are then obtained by using the simple relation
\begin{equation}
H_{obs}=-\dfrac{1}{(1+z)}{\left(\dfrac{dt}{dz}\right)}^{-1}\ .
\end{equation}
In \Cref{tab:OHD} of the Appendix we provide a list of 31 uncorrelated Hubble rate data over the interval $0<z<2$. The normalized Likelihood function in this case reads
\begin{equation}
\mathcal{L}_{OHD}=\dfrac{\exp\left[-\dfrac{1}{2}\displaystyle{\sum_{i=1}^{31}}\left(\dfrac{H_{th}(z_i)-H_{obs}(z_i)}{\sigma_{H,i}}\right)^2\right]} {\left[{(2\pi)}^{31}\displaystyle{\prod_{i=1}^{31}} \sigma_{H,i}^2\right]^{1/2}}\ .
\end{equation}

\subsection{Baryon acoustic oscillations}

The characteristic peaks in the galaxy correlation function are the imprint of baryon oscillations in the primordial plasma. A common procedure to quantify this physical phenomenon consists of estimating the combination of the comoving sound horizon at the drag epoch, $r_d$, and the spherically averaged distance $D_V(z)$ introduced in \cite{Eisenstein05}:
\begin{equation}
d_V^{th}(z)\equiv r_d\times D_V(z)^{-1}= r_d \left[\dfrac{{d_L}^2(z)}{(1+z)^2}\dfrac{z}{H(z)}\right]^{-1/3}.
\end{equation}
In this work, we use a collection of 6 model-independent BAO measurements presented in \cite{Lukovic16} and listed in \Cref{tab:BAO} of the Appendix. Since these are uncorrelated data, the normalized Likelihood function is given by
\begin{equation}
\mathcal{L}_{BAO}=\frac{\exp{\left[-\dfrac{1}{2}\displaystyle{\sum_{i=1}^6}\left(\dfrac{d_V^{th}(z_i)-d_V^{obs}(z_i)}{\sigma_{d_{V,i}}}\right)^2\right]}}{\left[{(2\pi)}^6\displaystyle{\prod_{i=1}^6}\sigma_{d_{V,i}}^2\right]^{1/2}}\ .
\end{equation}

\subsection{Numerical results and statistical model selection}

\begin{figure*}
\begin{center}
\includegraphics[width=1\textwidth]{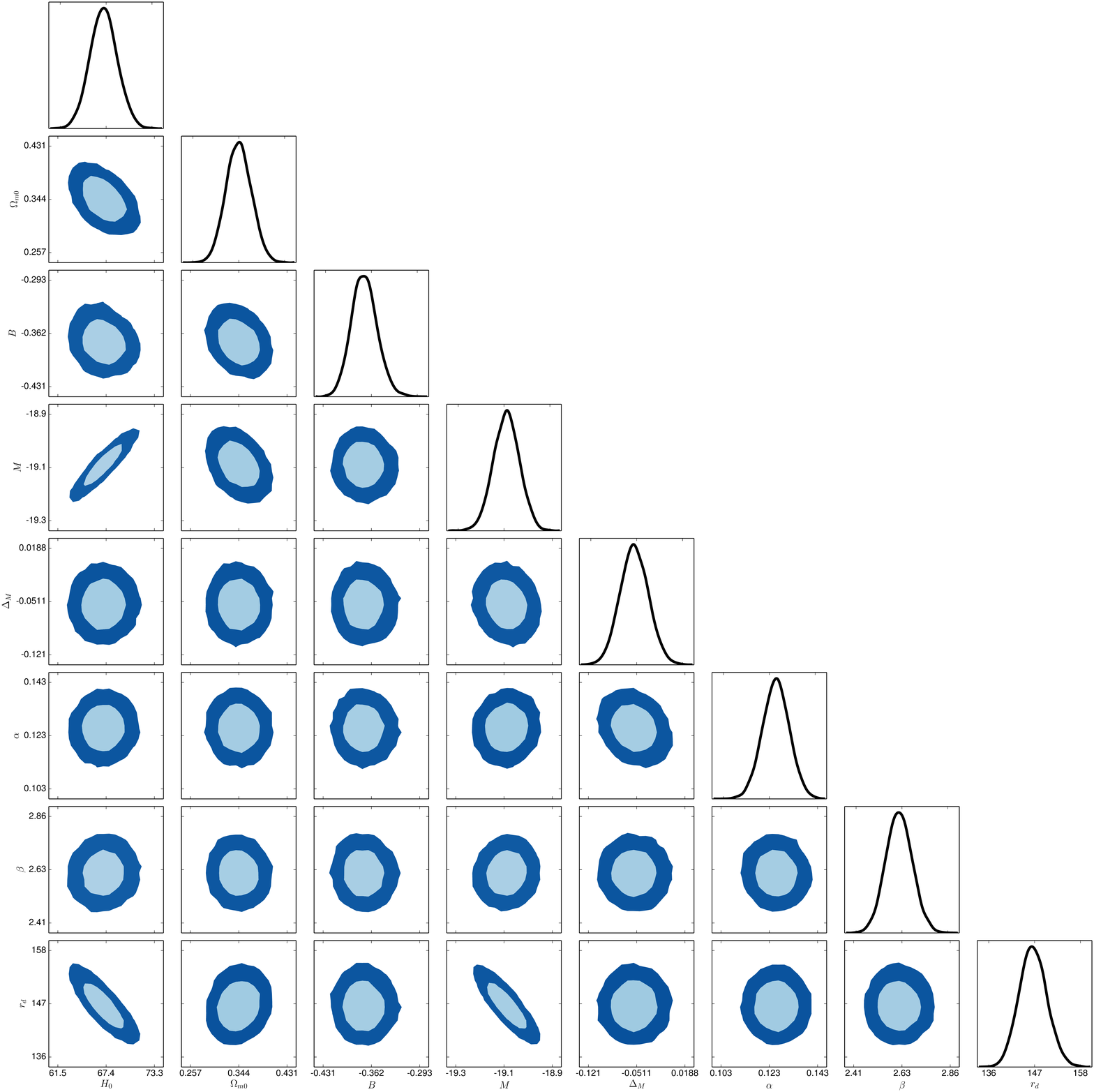}
\caption{2D marginalized 68\% and 95\% confidence levels contours, and 1D posterior distributions as result of the Monte Carlo numerical integration of the Anton-Schmidt model with $n=-1$.}
\label{fig:contours}
\end{center}
\end{figure*}

We perform Markov Chain Monte Carlo (MCMC) numerical integration through the Metropolis-Hasting algorithm implemented by the Monte Python code \cite{montepython}. The statistical analysis is done by considering the joint Likelihood of the combined data:
\begin{equation}
\mathcal{L}_{joint}=\mathcal{L}_{SN}\times \mathcal{L}_{OHD}\times \mathcal{L}_{BAO}\ .
\end{equation}
Uniform priors have been used for the cosmological as well as for the nuisance parameters. Our numerical results are presented in \Cref{tab:fit}, while in \Cref{fig:contours} we show the 2D $1\sigma$ and $2\sigma$ contours and the 1D posterior distributions.

\begin{table}[h!]
\begin{center}
\renewcommand{\arraystretch}{1.6}
\setlength{\tabcolsep}{1.5em}
\begin{tabular}{c c c }
\hline
\hline
Parameter & Prior & Result \\
\hline
$H_0$  & $(50,90)$ & $67.06^{+1.74}_{-1.85}$\\
$\Omega_{m,0}$ & $(0,1)$ & $0.344^{+0.024}_{-0.025}$ \\
$B$ & $(-1,0)$ & $-0.372^{+0.018}_{-0.021}$\\
$M$ & $(-20,-18)$ & $-19.08^{+0.06}_{-0.06}$\\
$\Delta_M$  & $(-1,1)$ & $-0.055^{+0.022}_{-0.022}$\\
$\alpha$  & $(0,1)$ & $0.126^{+0.006}_{-0.006}$\\
$\beta$  & $(0,5)$ & $2.618^{+0.066}_{-0.069}$\\
$r_d$  & $(130,160)$ & $146.7^{+3.3}_{-3.6}$ \\
\hline
\hline
\end{tabular}
\caption{Priors and 68\% confidence level parameter results of the MCMC analysis on the combined data for the Anton-Schmidt model with $n=-1$. $H_0$ and $r_d$ values are expressed in the usual units of km/s/Mpc and Mpc, respectively.}
 \label{tab:fit}
\end{center}
\end{table}

It is interesting to compare the results we have obtained for the Anton-Schmidt model  to the predictions of the standard $\Lambda$CDM model, whose expansion rate is given by
\begin{equation}
H_{\Lambda\text{CDM}}(z)=H_0\sqrt{\Omega_{m,0}(1+z)^3+\Omega_\Lambda}\ ,
\label{eq:H_LCDM}
\end{equation}
where $\Omega_{\Lambda}=1-\Omega_{m,0}$. In \Cref{tab:LCDM}, we show the results of the MCMC analysis for  $\Lambda$CDM.

\begin{table}[h!]
\begin{center}
\setlength{\tabcolsep}{1.5em}
\renewcommand{\arraystretch}{1.6}
\begin{tabular}{c c  }
\hline
\hline
Parameter & Result \\
\hline
$H_0$  & $66.56^{+1.24}_{-1.17}$ \\
$\Omega_{m,0}$  & $0.303^{+0.025}_{-0.023}$ \\
$M$ &  $-19.16^{+0.04}_{-0.04}$  \\
$\Delta_M$  & $-0.078^{+0.023}_{-0.019}$   \\
$\alpha$  & $0.122^{+0.006}_{-0.006}$ \\
$\beta$ & $2.570^{+0.069}_{-0.069}$ \\
$r_d$  & $146.1^{+1.9}_{-1.8}$\\
\hline
\hline
\end{tabular}
\caption{68\% confidence level results of the $\Lambda$CDM model from the MCMC analysis on the combined data. $H_0$ values are expressed in units of km/s/Mpc, and $r_d$ in units of Mpc.}
 \label{tab:LCDM}
\end{center}
\end{table}

A useful tool to select, among cosmological models, the one preferred by the data, is the Akaike information criterion (AIC) \cite{Akaike74} defined as
\begin{equation}
\text{AIC}=-2\ln \mathcal{L}_{max}+2p\ ,
\end{equation}
where $\mathcal{L}_{max}$ is the value of the Likelihood calculated for the best-fit parameters, and $p$ is the number of parameters in the model. The difference $\Delta\text{AIC}=\text{AIC}_i-\text{AIC}_j$ between the models $i$ and $j$ provides us with the best model, which corresponds to the one that minimizes the AIC value.
We also use the BIC criterion \cite{Schwarz78}:
\begin{equation}
\text{BIC}= -2\ln \mathcal{L}_{max}+p\ln N\ ,
\end{equation}
where $N$ is the number of the data.
In contrast with a simple comparison of the maximum Likelihood (or $\chi^2$ analysis), information criteria such as AIC or BIC compensate for any improvement in the maximum Likelihood that the introduction of extra parameters might allow. A more severe penalisation against the model with a larger number of free parameters is peculiar of the BIC criterion, due to the presence of the logarithm of the total number of data.  Here, we choose $\Lambda$CDM as the reference model, since it represents statistically the simplest cosmological model with the least number of parameters.
As shown in \Cref{tab:AIC-BIC}, both the selection criteria indicate a decisive evidence for the Anton-Schmidt model over $\Lambda$CDM.

\begin{table}[h!]
\begin{center}
\setlength{\tabcolsep}{1.2em}
\renewcommand{\arraystretch}{1.3}
\begin{tabular}{c c c c  }
\hline
\hline
Model & $\chi^2_{d.o.f.}$ & $\Delta$AIC & $\Delta$BIC \\
\hline
$\Lambda$CDM & $1.012$ & $0$ & 0 \\
Anton-Schmidt & $0.962$ & $-23.8$ & $-19.1$\\
\hline
\hline
\end{tabular}
\caption{$\chi^2$ per degree of freedom, AIC and BIC differences based on the best-fit results of the MCMC analysis for $\Lambda$CDM and the Anton-Schmidt model with $n=-1$.}
 \label{tab:AIC-BIC}
\end{center}
\end{table}

\subsection{Comparison with different cosmological models}

We compare here the predictions of the Anton-Schmidt model with different cosmological scenarios. In addition to $\Lambda$CDM (cf. \Cref{eq:H_LCDM}), we consider the $\omega$CDM model:
\begin{equation}
H_{\omega\text{CDM}}=H_0\sqrt{\Omega_{m,0}(1+z)^3+(1-\Omega_{m,0})(1+z)^{3(1+\omega)}}\ .
\label{eq:H_wcdm}
\end{equation}
According to the latest Planck collaboration results \cite{Planck15}, we have $\Omega_{m,0}=0.3065$ and $\omega=-1.006$.

Then, we also consider the dark energy parametrization represented by the Chevallier-Linder-Polarski (CPL) model \cite{CPL}:
\begin{align}
H_\text{CPL}&=H_0\left[\Omega_{m,0}(1+z)^3+(1-\Omega_{m,0})(1+z)^{3(1+\omega_0+\omega_1)}\right. \nonumber \\
&\hspace{1.1cm} \times \left. e^{-\frac{3\omega_1z}{1+z}}\right]^{1/2}.
\label{eq:H_CPL}
\end{align}
For this model, we adopt the values obtained by latest release of the WMAP project \cite{WMAP9}: $\Omega_{m,0}=0.2855$, $\omega_0=-1.17$ and $\omega_1=0.35$.

Finally, we consider the generalized Chaplygin gas (GCG) model \cite{Chaplygin}:
\begin{align}
H_\text{GCG}&=H_0\left\{\Omega_{m,0}(1+z)^3+(1-\Omega_{m,0})\Big[A_s+(1-A_s) \right.\nonumber \\
&\hspace{1.2cm}  \times  \left. (1+z)^{3(1+\alpha)}\Big]^{\frac{1}{1+\alpha}}\right\}^{1/2}\ ,
\end{align}
where $A_s$ is the present equation of state of the GCG fluid (see \cite{Chaplygin2} for the details).  The best-fit results found in \cite{Xu10} are: $\Omega_{m,0}=0.276$, $A_s=0.760$ and $\alpha=0.033$.

\begin{figure}[t]
\begin{center}
\includegraphics[width=3.2in]{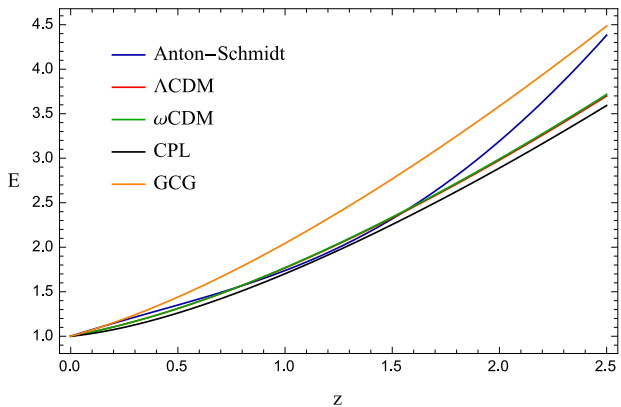}
\caption{Dimensionless expansion rate for different models.}
\label{fig:comparison}
\end{center}
\end{figure}

Using the outcomes of our MCMC analysis for the Anton-Schmidt and the $\Lambda$CDM models, and the values indicated above for the other cosmological scenarios, we show in \Cref{fig:comparison} the comparison of the different dimensionless expansion rates ($E(z)\equiv H(z)/H_0$) in the low-redshift regime.

\begin{figure}[t]
\begin{center}
\includegraphics[width=3.2in]{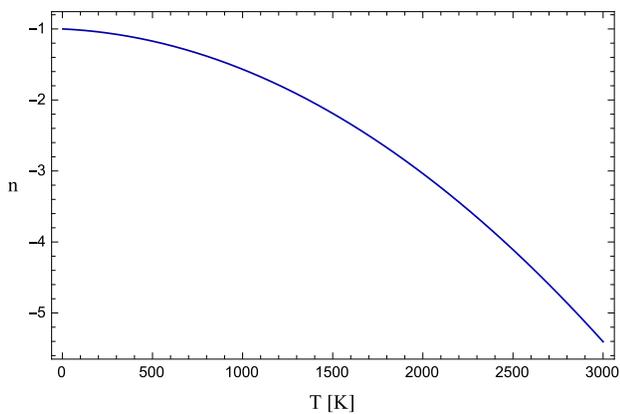}
\caption{Functional behaviour of the $n$ parameter of Anton-Schmidt's equation of state with temperature.}
\label{fig:n-T}
\end{center}
\end{figure}

\subsection{Comparison with Chaplygin gas}

The Anton-Schmidt approach candidates to unify dark energy and matter into a single dark fluid \cite{orlando}. The unifying dark energy models represent widely-appreciated scenarios whose main advantages lie on framing large-scale dynamics within a \emph{single fluid description} \cite{orlando2}. Among several possibilities, the Chaplygin gas represents a suitable prototype \cite{Chaplygin}, whose net pressure is under the form $P=-A/\epsilon$ with $A>0$.

The model has reached considerable success for its interconnection with string theory d-branes \cite{Chap2}. Unfortunately some experimental flaws plagued the model, leading to the formulation of a subsequently generalized Chaplygin gas, with pressure $P=-A/\epsilon^{\alpha}$ \cite{Chap3}. The extra term, namely $\alpha$ is bounded inside the interval $0\leq\alpha\leq1$ and provides a causal and subluminal sound speed, i.e. $0\leq c_s\leq1$.

One can recover the model through a k-essence Lagrangian, $\mathcal{L}\equiv\mathcal{L}\left(X\right)$, where $X = \frac{1}{2}\nabla_{\mu}\phi \nabla^{\mu}\phi$ and $\nabla_{\mu}$ the covariant derivative with respect to the coordinate $x^{\mu}$. 
Choosing as $\mathcal{L}\left(X\right)$ the following lagrangian density
\begin{equation}\label{beta BI Lagrangian}
\mathcal{L}=-\epsilon_\Lambda\sqrt{\left(1-\left(2X\right)^{\beta}\right)^{\frac{2\alpha}{1+\alpha}}}\,,
\end{equation}
with $0\leq 2X\leq 1$, one recovers the generalized version of the Chaplygin gas when $\beta=\left(1+\alpha\right)/(2\alpha)$. In this puzzle,  $\alpha$ and $\beta$ are positive constants whereas $\epsilon_\Lambda$ is a positive constant energy density.

The Anton-Schmidt paradigm can be related to the modified Chaplygin gas. In fact, the sound speed in a Chaplygin-like scenario is given as
\begin{equation}\label{GCG_c2}
c_{s}^{2}=\alpha\frac{A}{\epsilon^{1+\alpha}}\,,
\end{equation}
and following the work \cite{Ferreira18}, we have
\begin{equation}
P=-\frac{A}{\epsilon_*^{\alpha}}\left(\frac{\epsilon_{*}}{\epsilon}\right)^{\alpha}+C\,,
\end{equation}
where $A/\epsilon_*^{\alpha}=\epsilon_\Lambda$ and $C$ is a constant of integration. Expanding around $\alpha=0$ one obtains
\begin{equation}
P=\epsilon_\Lambda\left[-1 + \alpha\ln\left(\frac{\epsilon}{\epsilon_{*}}\right) + \mathcal{O}\left(\alpha^{2}\right)\right]+C\,.
\end{equation}
Taking into account the limit
\begin{equation}
\mathcal{A}=\lim_{\substack{\alpha\rightarrow 0\\ \epsilon_\Lambda\rightarrow\infty}} \alpha \epsilon_\Lambda\ ,
\end{equation}
with finite $\mathcal{A}$ and setting $C=\epsilon_\Lambda$, we have
\begin{equation}\label{logCG}
P=\mathcal{A}\ln\left(\frac{\epsilon}{\epsilon_*}\right).
\end{equation}
The model above defined is a version of  \emph{Logarithmic Chaplygin gas} and turns out to naively extend the  $\Lambda$CDM concordance model. The most intriguing fact is that the aforementioned EoS for Logarithmic Chaplygin gas corresponds to $n=0$ in the picture of Anton-Schmidt. This may be view as an alternative picture to get both the Anton-Schmidt fluid and logotropic versions of dark energy. Another intriguing fact would be the possibility to frame out a Lagrangian formalism associated to both the  logarithmic correction to the Chaplygin gas and to the Anton-Schmidt paradigm. Details are reported in \cite{Ferreira18,rocco}.

\section{Consequences at early times}
\label{sec:early}

The observational constraints obtained from the low-redhifts data can be used to infer the quantity $\rho_\ast$. In fact, inverting \Cref{eq:B} and using the best-fit results  of the MCMC analysis (cf. \Cref{tab:fit}) one gets
\begin{equation}
\dfrac{\rho_\ast}{\rho_{c,0}}=\Omega_{m,0}\ e^{-1/B}\simeq 5.1\ .
\end{equation}
This is contrast with the condition we have obtained from theoretical considerations at early times (see \Cref{subsec:early}) and with the limits found in \cite{Ferreira17}.
Moreover, we can check the consistency of our model with the observational constraints from the growth rate of linear perturbations  \cite{Ferreira17}. In the matter-dominated epoch, the comoving Jeans length is given by
\begin{equation}
\lambda_J^c=c_s(1+z)\sqrt{\dfrac{8\pi^2}{\rho}}\ .
\end{equation}
Using \Cref{eq:cs nuovo} and $\rho=3H_0^2\Omega_{m,0}(1+z)^3$, we find
\begin{equation}
\lambda_J^c=\dfrac{4\pi}{H_0\Omega_{m,0}}\sqrt{\dfrac{B(1-\Omega_{m,0})\left[1+B+3B\ln(1+z)\right]}{3(1+z)}}\ .
\end{equation}
This quantity defines the scale above which the linear growth of density perturbations can occur. Choosing $R=8h^{-1}$ Mpc,  the condition for the cosmic structures to grow in the matter era $(z>1)$ is $\lambda_J^c<R$. Such a condition is, however, not satisfied for the best-fit values of the cosmological parameters obtained from our numerical fit (cf. \Cref{tab:fit}).
The reason of aforementioned inconsistencies lies in the fact that Anton-Schmidt's equation of state is valid only in the Debye regime. Such an approximation is no longer licit as temperatures become high, making our model with $n=-1$ unpredictive at early times.
To solve this issue, one needs to consider the temperature dependence of the Gr\"uneisen parameter \cite{Tdep}.
Several theories have been proposed in the literature \cite{Tdep2}. A good agreement with experimental data is represented by the model proposed in \cite{Nie10}, according to which the Gr\"uneisen parameter can be expressed in terms of the temperature as follows:
\begin{equation}
\gamma_G(T)=\gamma_{G,0}\left[1+b_1(T-T_0)+b_2(T-T_0)^2\right] .
\label{eq:gamma_T}
\end{equation}
Here, $T_0$ is a reference temperature \footnote{In the case of solids, $T_0$ is the room temperature: 300 K.}, while $b_1$ and $b_2$ are free coefficients obtained by fitting experimental data at different temperatures. The authors in \cite{Nie10} found $b_1=1.45\times 10^{-4}$ K$^{-1}$ and $b_2=5.40\times 10^{-7}$ K$^{-2}$. Due to \Cref{eq:gamma_T}, the $n$ parameter of Anton-Schmidt's equation of state becomes also temperature-dependent:
\begin{equation}
n(T)=-\dfrac{1}{6}-\gamma_G(T)\ .
\label{eq:n_T}
\end{equation}
The above relation can be thus used to calibrate the value of $n$ at any temperature. In our case, we can identify $T_0$ with the temperature of the Cosmic Microwave Background (CMB) radiation today, $T_0=2.726$ K \cite{Fixsen09}.
We also fix $\gamma_{G,0}$ to the value correspondent to $n=-1$, so that $\gamma_{G,0}=5/6$. \Cref{fig:n-T} shows the values of $n$ at different temperatures.
As an example, at the last scattering surface, when the temperature of the CMB was $T=3000$ K, the value of $n$ reads
\begin{equation}
n_\text{CMB}=-5.40\ .
\end{equation}
More detailed analyses will be subject of future works.


\section{Outlook and perspectives}
\label{sec:conclusion}

In the present work, we propose a new class of dark energy models based on the assumption that matter obeys Anton-Schmidt's equation of state. In particular, we considered a single fluid description based on the use of scalar field with a given potential. We showed that Anton-Schmidt's pressure naturally provides a negative value even for matter only. This turns out to be true as one assumes that the cosmic expansion changes the thermodynamics of standard matter. Indeed, we demonstrated that, under the Debye approximation, one recovers a negative pressure proportional to the matter density itself, $\rho$. Relating it to the volume and to the field $\varphi$, we were able to frame the cosmic dynamics choosing the case $n=-1$ compatible with observations. In such a case, we found an analytic solution for $\varphi$, $\rho$ and the effective barotropic factor, $\omega$.
Through analyses of different regimes characterized by different values of the scale factor, we found that pressure vanishes at early regimes, while becomes significantly negative at late times.

We investigated the features of the new model at the level of background cosmology. 
We thus computed the adiabatic sound speed and related it to the case of pure logotropic model. In doing so, we pointed out the differences and the limits of our approach with respect to logotropic fluid scenarios.

Furthermore, we employed cosmological data such as Supernovae Ia, Hubble rate data and baryon acoustic oscillations to get observational constraints on our model. In particular, the Monte Carlo analysis on the combined data showed that our model statistically performs even better than the standard $\Lambda$CDM model. We thus discussed the consequences of our outcomes and we found that a generalization of the model is necessary to accommodate the constraints from the high-redshift regimes. Specifically, we proposed a model based on a temperature-dependent Gr\"uneisen parameter which would be able to satisfy the theoretical predictions of early time cosmology.

Future investigations will be dedicated to the analysis of the CMB data that will be used to calibrate the background evolution at early times  and match the results of low-redshift observations in order to remove possible degenerations (see also \cite{ananda} for a discussion on this topic).

\begin{acknowledgements}
This paper is based upon work from COST action CA15117  (CANTATA), supported by COST (European Cooperation in Science and Technology). S.C. acknowledges the support of INFN (iniziativa specifica QGSKY). 
The authors are very grateful to the anonymous referee for interesting requests and suggestions.
\end{acknowledgements}

\appendix*
\section{Experimental data}

\begin{table}[h]
\small
\begin{center}
\setlength{\tabcolsep}{1.5em}
\begin{tabular}{c c c }
\hline
\hline
 $z$ &$H \pm \sigma_H$ &  Reference \\
\hline
0.0708	& $69.00 \pm 19.68$ & \cite{Zhang14} \\
0.09	& $69.0 \pm 12.0$ & \cite{Jimenez02} \\
0.12	& $68.6 \pm 26.2$ & \cite{Zhang14} \\
0.17	& $83.0 \pm 8.0$ & \cite{Simon05} \\
0.179 & $75.0 \pm	4.0$ & \cite{Moresco12} \\
0.199 & $75.0	\pm 5.0$ & \cite{Moresco12} \\
0.20 &$72.9 \pm 29.6$ & \cite{Zhang14} \\
0.27	& $77.0 \pm 14.0$ & \cite{Simon05} \\
0.28	& $88.8 \pm 36.6$ & \cite{Zhang14} \\
0.35	& $82.1 \pm 4.85$ & \cite{Chuang12}\\
0.352 & $83.0	\pm 14.0$ & \cite{Moresco16} \\
0.3802	& $83.0 \pm 13.5$ & \cite{Moresco16}\\
0.4 & $95.0	\pm 17.0$ & \cite{Simon05} \\
0.4004	& $77.0 \pm 10.2$ & \cite{Moresco16} \\
0.4247	& $87.1 \pm 11.2$  & \cite{Moresco16} \\
0.4497 &	$92.8 \pm 12.9$ & \cite{Moresco16}\\
0.4783	 & $80.9 \pm 9.0$ & \cite{Moresco16} \\
0.48	& $97.0 \pm 62.0$ & \cite{Stern10} \\
0.593 & $104.0 \pm 13.0$ & \cite{Moresco12} \\
0.68	& $92.0 \pm 8.0$ & \cite{Moresco12} \\
0.781 & $105.0 \pm 12.0$ & \cite{Moresco12} \\
0.875 & $125.0 \pm 17.0 $ & \cite{Moresco12} \\
0.88	& $90.0 \pm 40.0$ & \cite{Stern10} \\
0.9 & $117.0 \pm 23.0$ & \cite{Simon05} \\
1.037 & $154.0 \pm 20.0$ & \cite{Moresco12} \\
1.3 & $168.0 \pm 17.0$ & \cite{Simon05} \\
1.363 & $160.0 \pm 33.6$ & \cite{Moresco15} \\
1.43	& $177.0 \pm18.0$ & \cite{Simon05} \\
1.53	& $140.0	\pm 14.0$ & \cite{Simon05} \\
1.75	 & $202.0 \pm 40.0$ & \cite{Simon05} \\
1.965& $186.5 \pm 50.4$ & \cite{Moresco15} \\
\hline
\hline
\end{tabular}
\caption{Observational $H(z)$ data in units of km/s/Mpc.}
 \label{tab:OHD}
\end{center}
\end{table}

\begin{table}[!h]
\begin{center}
\setlength{\tabcolsep}{1.5em}
\begin{tabular}{ c c c c }
\hline
\hline
 $z$ &$d_V \pm \sigma_{d_V}$ &  Reference \\
\hline
0.106 & 0.336 $\pm$  0.015   &  \cite{Beutler11}\\
0.15 & 0.2239 $\pm$ 0.0084  & \cite{Ross15} \\
0.32 &  0.1181 $\pm$ 0.0023 & \cite{Anderson14} \\
0.57 & 0.0726 $\pm$ 0.0007  & \cite{Anderson14} \\
2.34 & 0.0320 $\pm$ 0.0016 &  \cite{Delubac15} \\
2.36 &  0.0329 $\pm$ 0.0012  & \cite{Font-Ribera14}\\
\hline
\hline
\end{tabular}
\caption{Baryon acoustic oscillations measurements.}
 \label{tab:BAO}
\end{center}
\end{table}



\end{document}